# A streamlined approach to particle ALD coating of spherical beads

Ryan P. Badman[a], Xinwei Wu[b], *Vincent J. Genova[c]

[a] Department of Physics, 109 Clark Hall, Cornell University, Ithaca, NY, 14850, rpb226@cornell.edu
[b] Cornell NanoScale Science and Technology Facility, 250 Duffield Hall, Cornell University, Ithaca, NY, 14850, wu@cnf.cornell.edu
[c] Cornell NanoScale Science and Technology Facility, 250 Duffield Hall, Cornell University, Ithaca, NY, 14850, genova@cnf.cornell.edu,
*Principal Investigator

## I. INTRODUCTION

Functionalized nanoparticles play an increasingly important role in biomedical sciences, materials science, biophysics and numerous other disciplines. The ability to coat nanoparticles with a conformal metal or dielectric cladding, to nanometer precision, is often highly desired for applications such as fabricating anti-reflective coatings for high refractive index nanoparticles[1], creating diffusion barriers[2,3], tuning catalysis[4], or masking a toxic core with a biocompatible shell[5]. Coating alumina on highly UV-absorbing, but potentially bio-damaging, oxidative titania nano- or micro- particles is one high impact use of the particles and coatings, as only 6 nm of alumina can quench the photocatalytic activities of $TiO_2$[6,7]. Since photocatalysis of $TiO_2$ can damage DNA or cells, quenching photocatalytic materials like $TiO_2$ can be important in applications from cosmetics to optical tweezer biophysics experiments.

In this note, we describe a simple, low-cost and time-efficient method to conformally coat nanoparticles with nanometer precision using atomic layer deposition (ALD) of alumina, titania, and platinum films. The nanoparticles are chemically synthesized monodisperse titania and silica nanospheres that have tunable size in either material and tunable refractive index for titania. The coating process is performed without the long particle soaking ALD steps[8], fluidized beds[9,10,11,12] or vibration or multi-component rotary stages commonly used in current literature[12,13,14]. By eliminating the long precursor soak times used in industry protocols for high surface area samples, we were able to reduce the time needed to thermally deposit 45 nm of alumina, for example, from 48 hours to 8 hours. Additionally, our shortened ALD recipe gives the same deposition rate on Si wafers as on nanospheres, so a silicon process monitor piece can be inserted into the ALD chamber simultaneously with the nanoparticles as a rate monitor, thus eliminating the need for costly transmission electron microscopy (TEM) to check the particle cladding thickness. The limitation of this process, mainly the scalability beyond a few grams of substrate powder, makes our recipe most applicable for academic research labs.

## II. EXPERIMENTAL METHODOLOGY

### A. Titania Bead Synthesis

High quality amorphous titania monodisperse nanospheres were grown following the protocol outlined in previous literature[15,16] (Figure 1), with some small changes. For example, we reduced the wait period from the recommended 12-24 hours after mixing the glycolated titanium IV butoxide precursor solution with the solution of acetone, water and Tween 20 to 20 minutes. We determined that the synthesis reaction still was complete within this shortened period. We chose the molarity of titania butoxide in acetone that corresponded to monodisperse 300-400 nm diameter titania bead formation. Dehydration by annealing at 500-600 C further improved the size uniformity of beads, as described in literature[15,16].

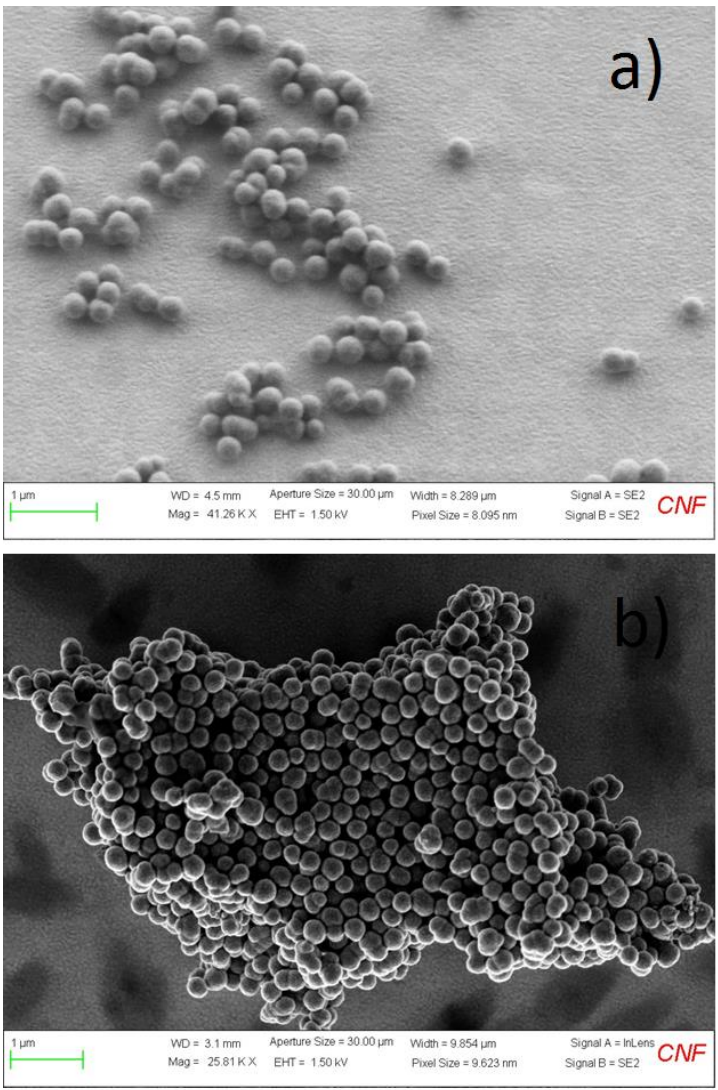

*FIG. 1. As-synthesized, uncoated mondisperse titania nanospheres images in an a) SE2 angled detector and b) Inlens detector in an SEM. Scale bars are 1 um in each figure.*

The solution of beads was so dense that the solution turned from clear to a milky white color after titania seed growth. However, perhaps either due to the presence of Tween 20 detergent or an effective de-clumping sonication and vortex protocol, aggregation only weakly occurred and did not hinder the ALD process as in other nanoparticle ALD reports[14]. After brief sonication and vortexing, and then dilution to typical optical tweezer trapping of bead concentrations, problematic clumping was not immediately rampant in aqueous or ethanol solution under a light microscope, as beads would typically float as single or double beads. However, we observed as expected from Derjaguin-Landau-Verwey-Overbeak (DLVO) and zeta potential theory, that higher index of refraction particles like titania clump more readily over time in solution than low index polymer or silica particles[17]. Thus storage of titania beads should be performed with proper steric[18] or ionic stabilization[19].

After three rounds of centrifugation and rinsing with ethanol, the newly grown titania beads were suspended in ethanol and then the dense solution was dried in an aluminum crucible approximately 2 inches in diameter or on a standard 100 mm Si wafer for deposition in the Arradiance ALD GEMStar6 system[20]. An optional step at this point is to anneal the beads in an atmospheric furnace at 500-600 C to create anatase phase titania, or 700-900 C to create rutile phase titania. Previous studies have shown that amorphous titania nanospheres have a refractive index of n=1.7-1.8, while annealed anatase and annealed rutile titania have refractive indices of 2.3 and 2.7 respectively in beads[1,15,16] (lower than the respective wafer titania thin films). We performed a 500 C anneal on the amorphous beads and observed a slight, permanent shrinkage of bead diameter from 350+/-70 nm down to 300 +/- 25 nm. The beads did not re-expand after suspension in solution, confirming previous results that the post-anneal size is highly stable[15]. We confirmed that the 500 C anneal produced an index higher than 2.0 using an optical trap experiment. Note that earlier work [21] found that alcohol rinses of synthesized titania can drastically lower the anatase-rutile phase transition temperature of titania nanoparticles. It should be mentioned that the anneal ambient conditions may be critical for the photocatalytic (and thus biocompatibility) properties of titania spheres. Literature suggests an oxygen or atmospheric anneal may be most desirable for low photocatalytic properties, while anneals in argon or nitrogen leave more oxygen vacancies and thus larger photocatalytic effects. The magnitude of the photocatalytic ability is inversely proportional the photoluminescent emission levels[22].

## B. Silica Bead Synthesis

We synthesized highly spherical silica beads using the standard ammonia and TEOS chemistry. The silica bead production was done according to modified version of previous work[23,24] to have a diameter comparable to the titania nanospheres to test the titania ALD process (Figure 2). The final silica bead diameters came out to be 375-550 nm. Our modifications allowed the beads to be produced in approximately one hour by mixing the reagents. A slightly longer and more complex process detailed in the literature[23,24] would produce silica beads of lower size variation of tunable size.

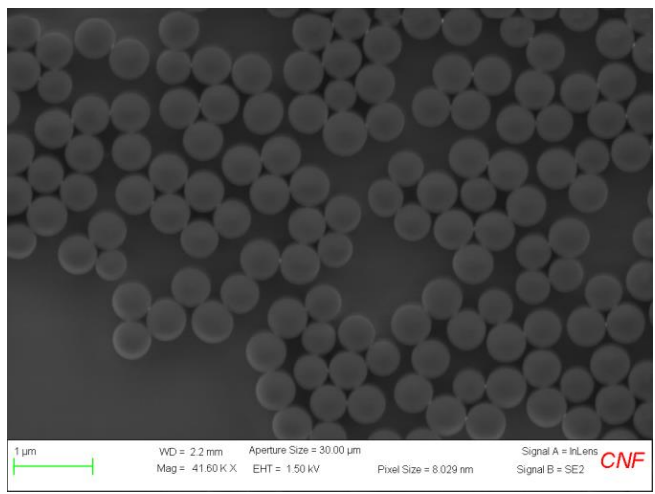

*FIG. 2. As-synthesized, uncoated, monodisperse silica nanospheres imaged in an Inlens SEM detector. Scale bar is 1 um.*

## C.  ALD Process Parameters

The ALD films were thermally deposited in the Arradiance GEMStar6 desktop system. The alumina films were deposited using trimethylaluminum (TMA) and water as the precursor for oxidation. Titania films were deposited by a reaction of tetrakis(dimethylamino)titanium(IV) (TDMAT) and water. Platinum films were deposited by a reaction of (trimethyl)methylcyclopentadienylPt(IV) Pt(MeCp)Me$_3$ and oxygen. For proper nucleation of Pt, a 1 nm seed layer deposition of alumina is incorporated in the recipe.

We next compared the recommended particle ALD recipe from Arradiance[8] to the planar ALD recipe optimized for rapid thermal ALD film growth on a wafer. The particle alumina ALD recipe ran for 120 cycles at 175 C with a 48 hour run time, while the alumina wafer growth recipe ran for 400 cycles at 175 C with an 8 hour run time. The process parameters for the particle ALD recipe are considerably different than the planar

ALD recipe to account for the extremely large increase in surface area needed to be conformally coated in beads vs a planar surface. The table below illustrates the significant differences in the $Al_2O_3$ particle and wafer ALD recipes. The cycle time of the particle alumina ALD recipe is 13.3min compared to only 0.7min of the wafer ALD recipe. To account for the much larger surface area, the particle ALD recipe incorporates 33 doses of TMA and 33 doses of water within a given ALD cycle. In addition, there are various delays in MFC flow stabilization and exposure valve operation within the particle recipe. A fast Pt wafer recipe deposition at 175 C was also applied to the titania nanospheres, and a 175 C titania wafer recipe to silica beads.

|  | Particle ALD | Wafer ALD |
|---|---|---|
| *Alumina ALD Step* | *Parameter Value* | *Parameter Value* |
| TMA Dose (msec) | 80 | 25 |
| Delay-hold (sec) | 12 | 18 |
| TMA doses/cycle | 33 | 1 |
| $H_2O$ dose time (msec) | 80 | 25 |
| Delay-hold (sec) | 12 | 23 |
| $H_2O$ doses/cycle | 33 | 1 |

*TABLE I. The alumina ALD coating steps and their parameter values for both the fast wafer coating recipe and the slow particle soaking recipe.*

## III. RESULTS AND DISCUSSION

Pre ALD-coated bead quality was studied using Zeiss Ultra and Zeiss Supra Scanning Electron Microscopes (SEMs), see figure 1 demonstrating the highly monodisperse titania spheres. An FEI T12 Spirit Transmission Electron Microscope (TEM) was used to image the post-ALD deposition coatings for many tens of beads per sample. A Woollam Ellipsometer was utilized to measure the ALD film thickness on a Si wafer placed in the ALD chamber during each bead run. Figure 3a demonstrates an alumina coating on a titania bead using the slow particle soaking recipe, compared to Figure 3b showing the same thickness alumina coating created in a high rate process for wafers that is six times faster than the particle soaking recipe. The coatings are virtually identical in quality and thickness in each recipe for the alumina ALD process. The coatings are also highly similar to those obtained using vibrating and rotary stages in ALD systems.

We verified that the high rate process produced highly conformal coatings in alumina of 15 nm, 30 nm, and 45 nm thicknesses. Additionally, we measured a wafer present in the chamber during the same deposition and found that the nanospheres' cladding thickness matched the wafer thickness to within 5 nm for titania beads with an alumina coating (Figure 3a and 3b).

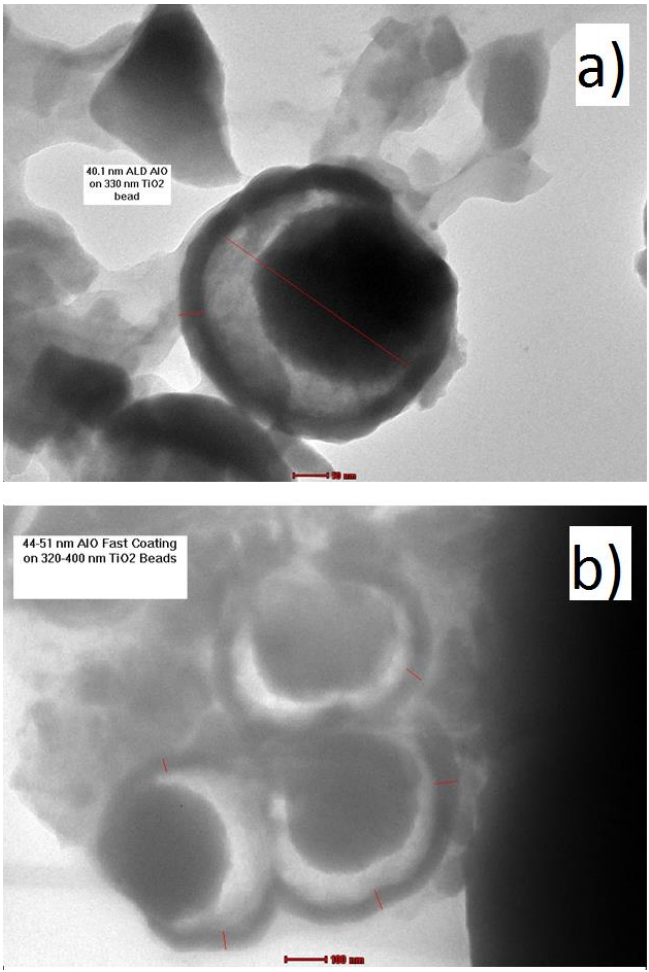

*FIG. 3. Alumina coated mondisperse titania nanospheres imaged in a TEM. The black shell on the outside of the beads is the alumina coating, while the less dense bead core is titania. Figure 3a depicts the results of the particle soaking recipe with a target of 40 nm to compare to 3b which has 45 nm of the faster ALD wafer recipe. The results look nearly identical. Scale bar is 50 nm in Figure 3a and 100 nm in Figure 3b.*

For our ALD Pt run on titania particles, we obtained 7-13 nm wide Pt grains covering the surface of approximately 200 nm titania nanospheres using a recipe that yielded 12 nm Pt on a wafer (Figure 4). Note that the metal ALD gives discrete grains while the dielectric coatings are more conformal, agreeing with previous literature[4].

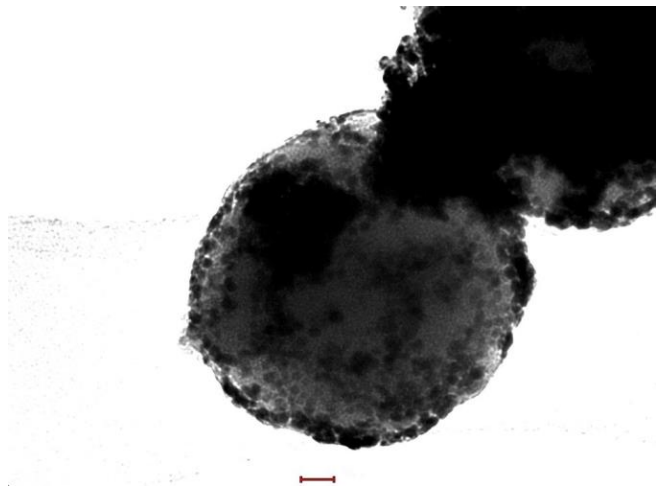

*FIG. 4. ALD Pt-coated titania nanospheres imaged in a TEM. The Pt coating is made up of 7-13 nm grains, colored black in the image. The grain thickness matches an ellipsometer measurement of a silicon monitor wafer coated with Pt in the same run as the titania spheres. Scale bar is 20 nm.*

For the monodisperse silica beads, the titania coating provided a 15-30 nm thick layer around the silica beads for a 30 nm wafer deposition (Figure 5). The ALD titania process was found to be slightly less uniform than the alumina ALD process. Each silica bead locally had a titania ALD coating that was conformal in thickness to approximately 5 nm, but across the bead population the bead coating thickness could vary between 15-30 nm.

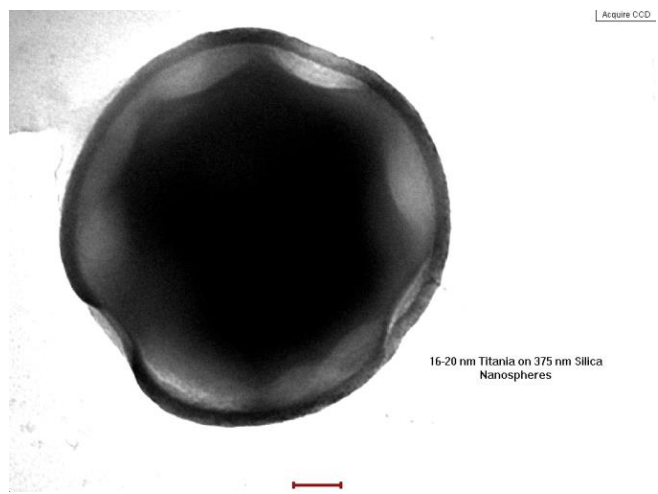

*FIG. 5. An example of a titania-coated silica nanosphere imaged in a TEM after a fast wafer ALD titania recipe. The (black) titania coating ranged from 15-30 nm on (gray) silica spheres for a run where the silicon monitor wafer was coated in 30 nm of titania. Scale bar is 50 nm.*

At a higher thickness of 85 or 100 nm ALD alumina wafer recipes, the conformal coating abilities were not uniform across the bead population in the crucible. The majority of beads had coating thicknesses varying between approximately 30-100 nm if this thicker recipe is attempted. Some of the variation in the smaller ALD thicknesses <40-50 nm, as well as the larger thicknesses, is likely an artifact due to particles not being perfectly spherical, so that the particle's tilt in the TEM makes the coating on one side of the particle look thicker than the other side[25]. One possible mechanism why a fast, static deposition is successful at lower thicknesses is that at this bead size and closed-packed spacing, the beads have enough space and thermal energy at 175 C to allow for ALD vapor penetration and uniform growth.

For larger amounts of bead powder added to the crucible, and with more tightly packed <50 nm particles, or coatings thicker than 40-50 nm, the conformal and uniform process is known to break down, and alternative methods become desirable. Fluidized beds, for example, offer a high precision route for scalable production of nanostructured particles[10,11], and allow for more efficient (up to 90% efficiency[12]) use of the precursor chemicals than static depositions. However, within our bead property parameter window, our runs used on the order of 1-10 milligrams per cm$^2$ of nanospheres in the stationary crucible, which is an amount we verified can last several weeks or longer in typical academic research labs, depending on the application. If a 100 mm or 150 mm diameter wafer is used to hold the powder, the quantity of beads can easily reach into the several grams range as the surface area for a standard 100 mm and 150 mm Si wafer is approximately 80 cm$^2$ and 176 cm$^2$ respectively.

For an excellent review further discussing the theoretical and experimental limitations of each ALD technique, including the fixed bed method explored in this report, see Longrie et al[26].

## IV. SUMMARY AND CONCLUSIONS

We conclude that time-consuming, expensive ALD processes with complicated sample holders may not be necessarily required for producing limited, but useful, quantities of highly conformal ALD coatings on nanoparticles. While planar stationary ALD deposition is not itself novel, we wanted to demonstrate and characterize the combination of streamlined and time-efficient nanosphere synthesis with ALD deposition for researchers entering the field to show that more advanced ALD techniques are not required for many nanotechnology applications.

We have demonstrated a highly repeatable and efficient nanofabrication process that allows monodisperse nanospheres of tunable size and index of refraction to be fabricated from scratch, and then precisely and uniformly conformally coated with up to 45 nm of dielectric ALD cladding within an 8-10 hour workday. The bead fabrication process takes approximately one hour while the ALD can be 1-8 hours depending on material and thickness, with silica ALD deposition rates exceeding those of alumina for example.

The entire process, from fabrication of monodisperse nanospheres to conformal atomic layer coating was done using shared user academic facilities. The process is low cost: the price of the process being set only by the cost of materials for 10-100 mL or more of high density titania nanosphere solution being fabricated and the ALD tool time of several hours. TEM measurement can be minimized if a monitor wafer piece is inserted into the ALD chamber. The beads can be placed in a standard stationary aluminum crucible or on a Si wafer up to six inches in diameter, and conformal coatings are obtained of equal quality to those reported with vibrating or rotary stages with these parameters.

## ACKNOWLEDGMENTS

We would like to acknowledge the helpful technical discussions with Dr. Huazhi Li of Arradiance and D.M. Tennant of CNF. This work was performed (1) in part at the Cornell NanoScale Science and Technology Facility (CNF), a member of the National Nanotechnology Infrastructure Network, which is supported by the NSF grant (ECCS-1542081), (2) made use of the Nanobiotechnology Center shared research facilities at Cornell, as well as (3) utilized the Cornell Center for Materials Science Research (CCMR) Shared Facilities which are supported through the NSF MRSEC program (DMR-1120296).